# Dynamical Diagnosis and Solutions for Resilient Natural and Social Systems

Tatyana Kovalenko and Didier Sornette
ETH Zurich
Department of Management, Technology and Economics
Scheuchzerstrasse 7, CH-8092 Zurich, Switzerland
(tkovalenko@ethz.ch and dsornette@ethz.ch)

**Abstract:** The concept of resilience embodies the quest towards the ability to sustain shocks, to suffer from these shocks as little as possible, for the shortest time possible, and to recover with the full functionalities that existed before the perturbation. We propose an operation definition of resilience, seeing it as a measure of stress that is complementary to the risk measures. Emphasis is put on the distinction between stressors (the forces acting on the system) and stress (the internal reaction of the system to the stressors). This allows us to elaborate a classification of stress measures and of the possible responses to stressors. We emphasize the need for characterizing the goals of a given system, from which the process of resilience build-up can be defined. Distinguishing between exogenous versus endogenous sources of stress allows one to define the corresponding appropriate responses. The main ingredients towards resilience include (1) the need for continuous multi-variable measurement and diagnosis of endogenous instabilities, (2) diversification and heterogeneity, (3) decoupling, (4) incentives and motivations, and (5) last but not least the (obvious) role of individual strengths. Propositions for individual training towards resilience are articulated. The concept of "crisis flight simulators" is introduced to address the intrinsic human cognitive biases underlying the logic of failures and the illusion of control, based on the premise that it is only by "living" through scenarios and experiencing them that decision makers make progress. We also introduce the "time@risk" framework, whose goal is to provide continuous predictive updates on possible scenarios and their probabilistic weights, so that a culture of preparedness and adaptation be promoted. These concepts are presented towards building up personal resilience, resilient societies and resilient financial systems.

**TABLE OF CONTENT**

**1-INTRODUCTION**

Interesting systems are out-of-equilibrium and subjected to external influences. In biology, the only true equilibrium state is death. In contrast, living organisms are remarkable engines that use energy and matter to generate internal order and external entropy. Being coupled to some outside environment, any interesting biological or social systems is under the influence of fluxes, their fluctuations and trends as well as perturbations of various types. Under these exogenous influences, they organize endogenously, attempting to self-propagate, grow and invade all available niches. These systems attempt to stabilize, at least for a time, towards some sort of dynamical equilibrium or are managed to stay close to a desirable state. Nevertheless, numerous exogenous and endogenous stress-factors continuously destabilize these systems. An outstanding question, which is increasingly crucial to modern human societies, is how to ensure survivability, sustainability, resilience as well as promise of better well-being and happiness in the presence of the many present and future stress factors.

To address these questions, the originality of the present essay is to recognize the key role played by the concept of "stress", which is the reaction of a system to some factors that tends to perturb it from a reference state. The existence of stress leads to three possible types of characteristics for a system:
> (i) fragility (system is prone to disability of its functions or even to destruction),
> (ii) robustness or resilience (system is able to recover from not-too-large stresses), and
> (iii) adaptiveness and transformation, leading to phase changes, regime shifts, modified behaviors and even to drastic structural reorganizations such as in biological mutations.

In this framework, we examine in detail the claim that stress can be beneficial and show that it is subdued within the earlier and more general concept of "adaptive systems" according to which systems evolve endogenously in symbiosis with the so-called stressors. The other essential role of stress in the evolution of systems is to promote rare intermittent rapid speciations, such as in punctuated biological evolution. We show that the concept of "antifragility" recently introduced by N. Taleb describes the quality of some systems that are designed to profit from particular stressors that stress other systems and to which they are not sensitive themselves. But, these so-called "antifragile" systems also exhibit vulnerability with respect to other stressors that lie outside their tailored design. Many presented antifragile systems are also much less productive that their fragile or resilient counterparts, showing the importance of recognizing the defined objectives. Hence, we conclude that antifragility does not exist per se and that the concept is misleading.

The present essay provides a rigorous definition of stress in corresponding systems. We describe how to measure stress, how to delineate the possible responses to stressors and we spell out propositions towards more resilience and sustainability. We emphasize the need for specifying the goals of a given system, from which the process of resilience build-in can be defined. We distinguish between exogenous versus endogenous sources of stress, and delineate the corresponding appropriate responses. We outline the main ingredients of resilience in terms of (1) the need for continuous multi-variable measurements and diagnosis of endogenous instabilities, (2) diversification and heterogeneity, (3) decoupling, (4) incentives and motivations, and (5) last but not least the (obvious) role of individual strengths. In this respect, propositions for individual training towards resilience are articulated. The concept of "crisis flight simulators" is introduced to address the intrinsic human cognitive biases underlying the logic of failures and the illusion of control, based on the premise that it is only by "living" through scenarios and experiencing them that decision makers make progress. We also introduce the "time@risk" framework, whose goal is to provide continuous predictive updates on possible scenarios and their probabilistic weights, so that a culture of preparedness and adaptation be promoted. These concepts are presented towards building up personal resilience, resilient societies and resilient financial systems.

**2-DEFINITIONS OF STRESS**

Defining stress is the first step towards a full understanding of risks, fragility, robustness, resilience and the development of efficient risk management. Everybody has some familiarity with the notion of stress. However, in view of the widespread misunderstanding and confusion, rigorous and precise definitions are required. Before formulating a general definition of stress, it is useful to present illustrations through examples offered by different scientific fields.

In physics and more specifically, in continuum mechanics, stress is defined as a measure of the internal forces acting within a deformable body. Quantitatively, we speak of a stress field defined as the ensemble of the stresses defined over all points within the body. Precisely, the stress at one point is a tensor that allows one to determine the force per unit surface that applies on any arbitrary fictitious plane specified by its orientation and going through that point. In a simple cylindrical geometry, an external force applied along the long axis translates within the body into a stress equal to the force divided by the area of the cylindrical cross-section. In equilibrium, the internal stresses sum up to balance exactly the external forces applied to the system. One can state the general result that the internal forces (and therefore stresses) are a reaction to external forces (stressors) applied on the body.

In biology, the endocrinologist pioneer, Hans Selye, introduced the concept of stress on the basis of his observations that many different types of substances and, more generally, perturbations applied to animals led to the same symptoms (see e.g., the review of Selye (1973)). The concept of stress in biology is thus based on the existence of non-specific responses of the body to the demands placed upon it. Transient perturbations, which do not exceed the natural regulatory capacity of the organism, lead to responses that ensure the resilience of homeostasis, the dynamical equilibrium characterizing living entities. In the presence of unrelieved stress, the body often transitions to pathological states associated with a change of homeostasis. This is analogous to the initial visco-elasto-plastic response of a mechanical system to an external stress, followed by creep that usually ends in the tertiary rupture regime (Nechad et al., 2005).

Several important characteristics of stress can be learnt from these examples:

1. stress is an internal response/reaction of a system to a perturbation called stressor (or stress-factor);

2. a stressor is a demand applied to the body that requires its reaction and adaptation;

3. a stressor elicits a non-specific response regardless of the nature of the stress, and even whether the stressor has a positive or negative consequences in the long term.

More generally, for biological as well as socio-economical systems, the non-specific response or "symptoms of stress" to a new demand involves increased

>(i) attention;
>
>(ii) mobilization of resources;
>
>(iii) concentration on key areas;
>
>(iv) recovery or exhaustion of the adaptive response and transition to pathological or crisis states.

In adaptive immune systems, (i) T and B lymphocytes first recognize the dangers, then (ii) mobilize the generating centers of antibodies that (iii) are finally directed towards and concentrated at the loci of insult. In social systems, the three first steps of the non-specific responses are typical of military-type intervention to cope with internal or external threats. In psychology, the first step (i) is associated with alarm, the second and third steps with resistance and the fourth step with exhaustion, as classified within the so-called general adaptation syndrome (Selye, 1973). More specifically, professionals facing acute situations, such as competitive pilots, athletes, surgeons and so on, go through the three first steps during their transient stressful activities. In economics, the response to economic difficulties is associated with (i) the characterization of the symptoms (solvency problems, budget deficit, increase of debt), (ii) the identification of reserves through expense cuts and reengineering of business and risk management processes and (iii) the reallocation of resources on key business lines or subsidizing. These measures may lead (iv) to a stabilization, or to a transition to a new favorable economic regime catalyzed by economic reforms and innovations or to bankruptcies in the context of firms, or to a disruptive transition to a new political order in the context of nations.

**3-MEASURES OF STRESS IN SOCIAL SCIENCE**

In mechanics, direct measurements of stress within a system are often performed by observations of deformations of the body. In biology, the measurement of stress is obtained by observing the response of the biological processes to a stressor. More generally, in natural sciences, one often has the luxury of observing the stresses by their direct effects. In social sciences, the feedback loops as well as coupling mechanisms to exogenous factors are much less understood. As a consequence of the complexity of social systems, the quantification of the stress level is performed indirectly via probabilistic approaches that introduce metrics of risks and/or resilience. These indirect ways of stress measurement in social sciences may be at the origin of the confusion in dealing with the concept of stress, incorrectly interpreted not as an internal response of the system to stressors but as the source of difficulties faced by the system.

**3.1 Risk as measure of stress**

Formally, risk is defined as the triplet of

1. a <u>probability</u> when available, or a level of uncertainty, or in the worst situation the formulation of the ambiguity corresponding to ask the question on the possibility for the occurrence of certain stressors;
2. a <u>potential loss</u> quantifying the possible impacts of the stressor;
3. a <u>vulnerability</u> and related counter-measures and mitigation techniques, that specify how disruptive is the potential stressor to the system.

The two first properties characterize the external forces or stressors that may influence the system. Together with the third property, which is specific to the system, they control the overall losses that the stressor can bring to the system. As a consequence, risk is understood as the combination of these three characteristics of the potential stressor. Thus, risk is equal or proportional to the possible internal response of the system, and therefore is a proxy for the stress developing within the system.

The simplest response of a system to a normal stress is non-specific and non-directional, which is comparable with the biological concept of kinesis. More resilient systems need to develop targeted reactions to stress, which is analogous to taxis in biology, defined as a directional response of a system to a stimulus or stimulus gradient intensity. In this sense, "stress taxis" can be defined as a response that, in the end, tends to unload stress off the system. For example, bacteria are wonderfully evolved organisms that demonstrate incredibly high resilience by using taxis and their corresponding simple behavioral rules.

**3.2 Resilience as measure of stress**

Resilience comes into several levels. The first two levels of resilience can be conveniently classified by using the theory of dynamical systems.

<u>First level of resilience.</u> Resilience is often defined as the speed of return to equilibrium (or more generally to the attractor characterizing the system) following a perturbation (Pimm and Lawton, 1977). Technically, the first level of resilience is referred to as "engineering resilience", which is a local concept. Engineering resilience is described by a local analysis, in terms of the stability of the linearized dynamics in the neighborhood of the equilibrium point. Indeed, resilience in this sense refers first to the stability of the equilibrium state, which occurs when all Lyapunov exponents are negative. Then, the speed of return to the equilibrium point is controlled by the largest (negative) Lyapunov exponent (i.e., the smallest one in absolute value).

<u>Second level of resilence</u>. In contrast, "ecological resilience" encompasses and generalizes engineering resilience by referring to the non-local dynamics occurring within the basin of attraction of the equilibrium state, defined as the set of initial conditions of the system that converge to that equilibrium state. While engineering resilience is a local concept quantifying the response of the system to small perturbations, ecological resilience describes the fact that a system state will return to its initial equilibrium as long as the perturbations remain within the basin of attraction of the equilibrium point, thus embodying non-local finite size perturbations that can be as large as the size of the basin of attraction itself, but not larger.

Walker et al. (2004) review four main components of ecological resilience of a system in its capacity to absorb disturbance and reorganize itself in order to retain essentially the same function. Using the dynamical system analogy with attractors and their basins of attraction, these four components are:
  (i) latitude (controlled by the size of the basin of attraction),
  (ii) resistance (controlled by the height of the barriers between attractors),
  (iii) precariousness (controlled by the current position of the system within the basin of attraction),
  (iv) and panarchy (controlled by the way the attractor structure and its basin may change as a function of the scale of description through cross-scale interactions) (Gunderson and Holling, 2002).

Extending the so-called resilience triangle approach (private communication of Wolfgang Kroger, ETH Zurich), see e.g. (Bruneau et al., 2003; Chang and Shinozuka, 2004; Pant and Barker, 2012), one can simplify the picture offered by ecological resilience by introducing four variables characterizing the response of a system to an external shock. Considering the variable $W_0$ corresponding to a reference capacity, wealth or production level just before the shock, we define
  (i) the maximum loss $(1-\lambda)W_0$,
  (ii) a characteristic time $\tau_1$ of reaction to reach the bottom level $\lambda W_0$,
  (iii) the level $\Lambda W_0$ recovered
  (iv) after the characteristic recovery time $\tau_2$.
In this simplified formulation, the resilience of the system is captured by the quadruplet of parameters ($\lambda$, $\tau_1$, $\Lambda$, $\tau_2$). Note that $\Lambda$ could be larger than 1, corresponding to the situation where the shock has long-term beneficial effects by increasing the overall performance above the initial baseline $W_0$. Some systems may be characterize by $\Lambda$ being smaller than $\lambda$, in which case, after a first loss of performance over a first reaction time $\tau_1$, the system degrades further over a possibly different time scale $\tau_2$ to an even worse situation. We should also stress that the quadruplet ($\lambda$, $\tau_1$, $\Lambda$, $\tau_2$) may not be unique but depend on the severity and duration (as well as possibly other characteristics) of the shock, so as to reflect the nature and amplitude of possible cascades occurring within the system.

<u>Third level of resilience</u>. The concept of viability (Aubin, 1991; Deffuant and Gilbert, 2011) extends further the idea by focusing on the conditions that the system must obey to remain "viable", for instance functional or alive. These constraints may not in general map precisely onto the set of attractors of the dynamics or may not even be attainable by the natural evolution of the dynamics and therefore may require continuous external management and control.

<u>Fourth level of resilience</u>. The dynamical system analogy has however its limit if taken too rigidly, because it fails to account for the fact that many biological, ecological and social systems may actually adapt, evolve and even transform fundamentally under the influence of stressors (Walker et al., 2004). This requires the consideration of other levels of resilience, which takes into account the possibility for the system to adapt its constituents so as to influence its resilience. This may correspond to a deformation of the basin of attraction, the fusion of initially distinct basins and other topological transformations. More generally, the dynamical system may incorporate stochastic components, such as deterministic, quasi-periodic or even random deformations of the attractors due to the modulation of some control parameters, as long as the conditions of viability are respected. Then, the system keeps its identity, but in a broader sense, even redefining itself while still keeping its ability to cope with the stressors. Pushed to the extreme, the system may even transform itself into a completely different structure via its capacity to evolve, as described by the theory of complex adaptive systems (Holland, 1975; Kauffman, 1993).

These considerations underline that the concept of resilience is dependent on the time scale over which the stressors act. For short-lived disturbances compared with the characteristic time scales of reactions of the system, engineering and ecological resiliences are the relevant levels of description. At intermediate time scales, the issue of viability dominates, pushing for adaptation and redefinition of goals and processes. At the longest time scales, transformations may occur that are similar to natural selection and Darwinist evolution of species, seen as a transformation in response to changing geological and climatic conditions. In the context of man-made and social systems, Darwinist evolution is also relevant to understand the dynamics of human enterprises (Hannan and Freeman, 1977; Hite and Hesterly, 2001). Real life situations are likely to involve an interplay between a continuum of different time scales and thus between the different levels of resilience.

**3.3 Links between risk and resilience as complementary measures of stress**

To summarize, risk and resilience are two complementary revelations of stress. On the one hand, risk provides a measure of the nature and amplitude of stressors, present and future. As a consequence, from risk measurements, one can infer the possible level of stress that may develop within the system. On the other hand, resilience characterizes the internal stress response within the system, quantified by the capacity of a system to cope with stressors and remain essentially the same. In other words, resilience is the amount of stress that a system can bear without a considerable transformation.

Risk and resilience are inter-connected in another way through the concept of vulnerability. On the one hand, vulnerability is part of risk, as a quantification of the potential amount of losses that are specific to a given system. But this vulnerability depends on the structural and adaptive properties of the system that make it either more prone to losses or less vulnerable via better mitigation techniques. In this sense, vulnerability constructs a bridge between risk and resilience. The processes favoring resilience will tend to decrease vulnerability and vice-versa.

The duality of stress expression in risk versus resilience is also apparent in the different possible responses of the system to stressors. These responses can be classified into three main classes: (i) fight, (ii) fly and (iii) transform.
(i) "Fight" is the typical response under relatively small risk and large resilience, which are associated with "normal" stress developing within the system. The "fight" response can be characterized by negative feedback loops tending to stabilize the system around its previous state, such as in the homeostasis state of living biological entities.
(ii) In contrast, the "fly" response corresponds to systems where risks and resilience are at comparable levels so that there is significant hazard for the system. By avoidance strategies, or some adaptation without major transformations and/or improvement of management, resilience can be improved so that the stressors can be addressed in order to ensure the preservation of the system identity.
(iii) Finally, when risk is large and resilience is insufficient, "extreme" stress develops within the system. Other than its demise, its survival requires considerable transformations of the system itself via the activation of positive feedbacks that drives it towards a new state.

The rational response to the presence of risks (the potential stressors and corresponding stress of the system) would seem logically to strive for always increasing resilience (the stress that the system can bear). However, there is always a cost-benefit balance between two extremes, the laissez-faire attitude of no investment in resilience as one extreme, and extreme risk aversion leading to attempts to over-control at the other. Building up resilience requires indeed to increase reserves, develop excess capacity, construct alternative supply chains, ensure redundancy, as well as investing in continuous education and training. But modern optimizing firms and societies work with the just-in-time philosophy and the constraint of ever lowering costs. This is often an impediment for building up resilience, as many examples show (Sheffi, 2005). It is a general observation that management in social systems strives to optimize this cost-benefit conflict, however, with often limited or even disappointing results to show for. In contrast, it is remarkable that natural systems often tend to evolve, converge and operate close to states that exhibit such a balance. These states are referred to in the modern literature as "self-organized critical" (Bak, 1996) or "at the edge of chaos" (Kauffman, 1993). This describes the tendency for coupled entities that interact over many repetitive actions to function close to a bifurcation point separating states that are too stable, from other states that are too unstable. A typical example is the human brain, for which there is a growing consensus that it operates close to or even functions at a critical point (Chialvo, 2006; Levina et al., 2007; Meisel et al., 2012; Plenz, 2012), separating a sub-critical state from a super-critical one. In the critical state, the brain exhibits the largest possible reactivity to novel external stimuli while, at the same time, showing stability of memory and other functional properties. If the brain was in the subcritical state, it would learn less efficiently by being not malleable enough and would be too slow to react in crucial situations. If the brain was in the supercritical state, its neural network would fire too much and too often, oscillating between extreme activity and exhaustion. Such a pathological state is actually found in epileptic patients (Osorio et al., 2010). In natural and biological systems, there are in general strong negative feedback mechanisms to stabilize the system and poise it at an optimal point between costly increase of resilience and costly neglect of the looming risks (Scheffer, 2009). The balance corresponds to a merging of the two responses - "fight" and "fly" - so that the system may combine both negative feedback reactions as well as adaptation to remain at the "edge of chaos".

In social systems, there is a lot of lip service paid to the goal for managers and policy makers to obtain this kind of optimal state. Actually, there is often an illusion of control (Langer, 1975; Satinover and Sornette, 2007; 2011) that it is possible to remove most of the risks and obtain an ideal state of resilience. One argument for the insufficient resilience of social systems (Diamond, 2004) is that, due to their complexity, they have not had the time to evolve (Walker et al., 2004) by the forces of "natural selection". This may be a part of the truth. However, we note that, for some social systems such as financial markets, there is ample evidence of an absence of convergence towards a stable dynamics, but rather the existence of persistent cycles of bubbles and bursts (Kindleberger, 2005; Sornette, 2003), notwithstanding experiencing many crises that, one would surmise, would have enabled investors to learn and avoid the next one (Reinhart and Rogoff, 2011). One possible explanation can be found in the incentives of investors to maximize their return on short time scales, leading to recurrent instabilities (Minsky, 2008). More generally, in many social systems, there is the ubiquitous problem that the short-term incentives are often not aligned with the long-term ones. This is associated with hyperbolic discounting (Laibson, 1998), which describes the general exaggerated preference for smaller immediate rather than larger delayed gratifications. Similarly, the incentives at the individual agent level are often incompatible with those at the society level, leading to social dilemmas (Kerr, 1983). It is also associated with the so-called public good problem and the problem of fostering social cooperation in particular in the context of socio-ecological systems (Ostrom, 1990). The rest of this essay aims at characterizing the conditions for breaking these kinds of stalemate.

**4-CAN STRESS BE BENEFICIAL?**

When thinking about stress, a first attitude is to find ways of reducing it or, when not possible, of developing passive and/or active defenses. But, there is a growing recognition that moderate levels of stress may be actually beneficial, both for health and for performance (Weiten and Lloyd, 2005; Hosenpud and Greenberg, 2006; Ritsner, 2010; Contrada and Baum, 2010). Is stress really beneficial per se?

**4.1 System-stressor co-evolution under normal stress**

For passive systems, stress is in general destructive, as in creep of materials where microscopic tiny damage events accumulate and lead to global rupture. In contrast, active systems can detect stress and use it as a guiding signal on the way towards better fitness to novel conditions. Thus, random or intended stressors are usable for the
   (i) identification of the characteristics of stress by listening and analyzing reactions of the system to perturbations;
   (ii) measurement of stress: (a) risks (observation of event probabilities, losses, vulnerability of the system) and (b) resilience ("exploration" of the stability landscape characterized by its latitude, resistance, precariousness and panarchy);
   (iii) catalysis of learning, which promotes changes occurring through feedback mechanisms by adaptation towards better fitness under changing conditions, and of selection of specific features and implementation of contra-measures;
   (iv) excitation of the system readiness, maintaining an engaged, interested and concerned state (in the spirit of the Soviet Union pioneer's motto "Always Ready!").

In section 2, we identified that symptoms of stress in a system include attention, mobilization of resources, concentrations on key areas, and so on. This may be viewed as positive consequences of stress for the function of the system. But, these changes are actually occurring at a cost, in particular that of a loss of resilience because the allocation of resources to cope with the stressor makes the system more vulnerable to other stressors. Thus, the optimization to cope with a first stressor should not be seen necessarily as a benefit of the stress. In general, optimization processes and coping with stress (or strengthening resilience) should be disconnected.

We also need to mention the cases in which some stress can be caused by a "positive" stress-factor (termed "Eustress" by Selye (1973)). For example, an eustress could be an economic reform that, after a period of adaptation, would lead to increased economic growth. Or, an extraordinary good news (learning about the return of a lost one or winning a huge lottery sum) may induce strong stress in the person. Again, it is not stress itself that is beneficial. Stress is a signal of a change of conditions and is a "guide" on the way towards adaptation or transformation to better fit to the new conditions, so that a system can survive and benefit from them.

Many situations where stress is argued to be beneficial, which we are going to cover at least partially in the following, follow the same archetype in which the system under consideration has co-evolved with the stress. In other words, the system is within an environment in which stress is unavoidable. Stress seems to be beneficial simply because the raison d'être of the system or of some of its key properties is precisely to cope and live with the ambient stress. Therefore, it is almost a tautology to find that the system needs stress or benefits from stress because it becomes dysfunctional if one of its main inputs, stress, is absent. We can therefore state that stress, at least up to a certain level smaller than the system resilience, is part of the normal system function and we refer to this situation as "normal stress".

**4.2 Adapted systems co-evolved with their stressors**

In this section, we provide several examples illustrating the concept that so-called beneficial stress occurs when the system under consideration has co-evolved with the stress.

### 4.2.1 Mammal immune systems, bones and muscles

Biology and medicine have probably been the first disciplines to recognize the co-evolved nature of stressors and of the stresses that develop within living systems. The immune system of mammals, in particular, provides arguably the best example illustrating what could be referred to with perhaps some exaggeration as a symbiosis between stressors (antigens) and system (antibodies). We underline that the example of the immune system provides a particularly important illustration, since its main role is indeed to defend the organism against disruptive intrusions by pathogens, in particular, which would like to exploit the organism for their own propagation. Consider first other types of homeostasis control processes in which the target variables are kept in a narrow optimal range with small fluctuations. This describes the "stable" homeostasis control for the regulation of the amounts of water and minerals by osmoregulation in the kidneys, the removal of metabolic waste by excretory organs such as the kidneys and lungs, the regulation of body temperature, the regulation of blood glucose level by the liver and the insulin secreted by the pancreas, and so on. In contrast, "The (immune) system never settles down to a steady-state, but rather, constantly changes with local flare ups and storms, and with periods of relative quiescence" as quoted in (Perelson, 2002), and see also (Perelson and Weisbuch, 1997; Nelson and Perelson, 2002). These flares can be understood as transient nonlinear reactions to fluctuating exogenous stressors as well as to expressions of the internal stress states. A growing body of literature indeed suggests that the incessant "attacks" by antigens of many different forms have forced the immune system to develop continuing fight and adaptation processes to ensure the integrity of the body (see Sornette et al. (2009b) for a review and mathematical modeling). In this vein, the 'hygiene hypothesis' (Schaub et al. 2006) states that modern medicine and sanitation may give rise to an under-stimulated and subsequently overactive immune system that is responsible for high incidences of immune-related ailments such as allergy and autoimmune diseases. In this view, infections and unhygienic contact may confer protection against the development of allergic illnesses. For instance, Bollinger et al. (2007) suggested that the hygiene hypothesis may explain the increased rate of appendicitis (~6% incidence) in industrialized countries, in relation to the important immune-related function of the appendix. Sornette et al. (2009) concluded that, if the regulatory immune system was not continuously subjected to stressors, its adaptive component would decay in part and the defense would go down, thus letting the organism becoming vulnerable to future bursts of pathogen fluxes. They developed a mathematical model that demonstrates that the correct point of reference is not a microbe-free body (no stressors), but a highly dynamical homeostatic immune system within a homeostatic body under the impact of fluxes of pathogens and of other stressors (which include microorganisms such as bacteria, viruses, fungi, parasites, environmental load, over-work, overeating and other excesses, psychological and emotional factors such as anger, fear, sadness, and so on). The situation is analogous to the maintenance of healthy bones and muscles of a human being. For astronauts under zero-gravity (no weight stressor), loss of bone and muscle, cardiovascular deconditioning, loss of red blood cells and plasma, possible compromise of the immune system, and finally, an inappropriate interpretation of otolith system signals all occur, with no appropriate counter-measures yet known (Young, 1999). In other words, for bones and muscles, stress (in the real mechanical sense of the term!) is needed to avoid degenerescence and ensure appropriate strength in cases of need. In all these examples, stress is beneficial only because the systems are fundamentally defined in their aims and properties by their interactions with stressors. Biological evolution has weaved a complex network of interacting feedback loops that entangled fundamentally the systems with their stressors, making the later necessary for the normal function of the former.

### 4.2.2 Human cooperation, competition and risk taking

An enormous body of anthropological and ethnographic literature demonstrates that the level of cooperation between humans is exceptional both in quality and quantity (Henrich and Henrich, 2007), which explains the remarkable success of this single mammal species that nowadays controls a major part of the whole output of planet Earth (Steffen et al., 2004). However, the origin of this cooperation is still quoted as one of the 25 most compelling puzzles that science is facing today (Siegfried, 2005). Many mechanisms and contextual factors have been proposed to explain the remarkable level of pro-social behavior and cooperation between humans, such as kin selection, inclusive fitness, reciprocity, network reciprocity, group-level and multi-level selection, other-regarding preferences, relative income preferences, envy, inequality aversion and altruism (Axelrod and Hamilton, 1981). Two essential ingredients emerge: (i) the presence of differences in skills, contributions, rewards and retributions among group members and (ii) how perceptions and preferences drive human decisions and actions. In other words, not only exogenous stressors resulting from the environment such as predators but also within-group stressors have been found essential to promote cooperation. This has led to a significantly higher survival efficiency and larger fitness both for the group and for the individuals. Using agent-based models and analytical theory, Hetzer and Sornette (2011; 2012) in particular have shown that cooperation evolves at the level documented for humans only under two conditions: (i) agents exhibit disadvantageous inequity aversion, which is found to be evolutionary dominant and stable in a heterogeneous population of agents endowed initially only with purely self-regarding preferences; (ii) groups are "stressed" by random perturbations in the form of strangers migrating between co-evolving groups and who introduce different cooperation levels than those that would emerge from the group consensus in absence of the random perturbations. The underlying mechanism is related to the Parrondo effect describing situations where losing strategies or deleterious effects can combine to win (Harmer and Abbott, 2002; Abbott, 2002). Here, the random behavior is rooted in the exchange between groups and the asymmetry is inscribed in the punishment rule driven by disadvantageous inequity aversion. This constitutes a telling example illustrating that stressors have selected for enhanced cooperation via higher survival rates for groups and individuals. This became possible when cognitive abilities in our homo ancestors increased sufficiently to allow the exploitation of this new "resource" of enhanced cooperation beyond that observed for our primate cousins, again illustrating the co-evolution between stressors and system's abilities.

Another important characteristic of humans is that high male-male competition for reproductive success has been permeating the history of modern humans (200'000 years ago to recent times) and has contributed through gene-culture coevolution to create gender competitiveness-related differences. Favre and Sornette (2012) have recently introduced a simple agent-based model that explains the high level of male-male competition and risk taking as rooted in the unequal biological costs of reproduction between males and females. This cost asymmetry has promoted females' choosy selection of alpha-males who have better chance to propagate genes via the natural selection of the fittest (Baumeister, 2010; Ogas and Gaddami). This causes male-male competition and male's arm race for signaling their qualities, which takes the form of stronger risk-taking behavior (Diamond, 2002). This further cascades into higher male than female death rates through risky signaling and results in a smaller male than female effective breeding population, both because females select a subset of males for reproduction and because of male's higher death rate. Remarkably, this mechanism can be checked quantitatively through its prediction for the ratio of the Time To the Most Recent Common Ancestor (TMRCA) based on human mitochondrial DNA (mtDNA), i.e. female-to-female transmitted, which is estimated to be twice that based on the non-recombining part of the Y chromosome (NRY), i.e. male-to-male transmitted. It appears that we are all descended from males who were successful in a highly competitive context, while females were facing a much weaker female-female competition. Stresses have appeared endogenously in the human population as a response to the unequal biological costs of reproduction (itself a stressor), leading to males' arm race in risk taking (another set of stressors) and cascaded into extraordinary implications for the development of the human species and its conquer of the world (Baumeister, 2010). One can argue that the high level of risk taking of human males have been beneficial for mankind, through the exploration of unknown territories and the development of inventions, in the end making stressors, via enhanced risk-taking by males, the engine of progress. The causal flow "reproduction inequality => female strategy => male risk taking" of stressors can thus be seen as an intrinsic part of the making of mankind, providing another example of the entangled nature of the human system and its stressors, the later being beneficial on the long term as a result of their co-existence and co-evolution. Pushing this reasoning, one can thus conclude that being human is to use one's superior cognitive abilities to take risks beyond the biological laws that enslave other animals.

## 4.3 Change of regimes under extreme stress

Nature and human societies exhibit many cases in history and in recent times when stress surpasses the resilience level of the system. We refer to such response of the system as "extreme stress" because of dramatic consequences it may lead to. Sources of extreme stress can be tracked using the measures of stress that were described above - risk and resilience - and include:
- (i) extreme possible stressors that are characterized by low probability and/or huge losses, for example, very rare events of enormous impact or previously unknown events (black swans (Taleb, 2007));
- (ii) unbearable stress that the system is not capable of coping with, showing extreme vulnerabilities (for example, disfunction of critical systems) and/or zero resilience, when even a tiny perturbation can lead to a change of regime. Examples of such systems include those (1) optimized to the edge of maximum efficiency, such as the just-in-time Toyota supply chain and inventory management system and (2) close to a tipping point due to developed endogenous instabilities, leading to dragon-kings (Sornette and Ouillon, 2012).

In the worst cases, this leads to the death or demise of the corresponding organism or system, as for instance documented by J. Diamond (2004) for human societies. In other situations, the system evolves to another regime, in which different properties that were dormant come into play or novel ones are forced to evolve for the survival and success of the system.

The following two subsections examine a number of real life examples illustrating the occurrence of regime shifts and evolution under extreme stress.

### 4.3.1 Biological and other transitions

The existence of changes of states promoted by extreme conditions is perhaps best incarnated by biological evolution. Contrarily to the initial view held by Darwin that evolution is generally smooth and continuous, occurring by the cumulative effect of gradual transformations, the theory of punctuated equilibrium in evolutionary biology describes the evolution of species as a sequence of stable states punctuated by rare and rapid events of branching speciations occurring under the stresses resulting from climatic, geographic and other possible evolutionary stressors (Gould and Eldredge, 1993). Since its introduction (Eldredge and Gould, 1972), this theory has received strong empirical support (Gould, 2002; Lyne and Howe, 2007). It holds that most species exhibit little evolutionary change for most of their geological history, being adapted to their niches. But, something happens, such as an extreme disturbance, that pushes the species to branch into novel species, often with the demise or altogether change of the original species.

Many scientists view the abrupt changes occurring in the sequence of punctuated equilibria as due to catastrophic causes, such as the famous Chicxulub asteroid (Schulte et al., 2010) or enormous volcanic eruptions in the so-called Deccan trap epoch (Courtillot and McClinton, 2002), or both (Archibald et al., 2010) ending the reign of the mighty dinosaurs about 65 millions years ago. Starting with Bak and Sneppen (1993), others have argued for an endogenous origin, using the analogy with the concept of self-organized criticality (Bak and Paczuski, 1995; Bak, 1996; Jensen, 1998; Sornette, 2004, chapter 15). According to complex system theory, out-of-equilibrium slowly driven systems with threshold dynamics relax through a hierarchy of avalanches of all sizes. Accordingly, extreme events can also be endogenous.

The exogenous versus endogenous explanations may actually represent two complementary view points since, in reality, they are often entangled. Indeed, how can one assert with 100% confidence that a given extreme event is really due to an endogenous self-organization of the system, rather than to the response to an external shock? Most natural and social systems are indeed continuously subjected to external stimulations, noises, shocks, stress, forces and so on, which can widely vary in amplitude. It is thus not clear a priori if a given large event is due to a strong exogenous shock, to the internal dynamics of the system, or maybe to a combination of both. Sornette et al. have advanced the hypothesis that specific dynamical signatures of precursors occurring before and relaxations following extreme events lead to a classification of possible regimes and the possibility to resolve the endo-exo conundrum. This applies broadly to many complex systems (Sornette and Helmstetter, 2003; Sornette, 2005), for which it is fundamental to understand the relative importance of self-organization versus external forcing, as documented for financial shocks (Sornette et al., 2003), commercial sales (Sornette et al., 2004), and for the dynamics of fame of YouTube videos (Crane and Sornette, 2008). More generally, in addition to biological extinctions such as the Cretaceous/Tertiary KT boundary (meteorite versus extreme volcanic activity versus self-organized critical extinction cascades), this question applies to commercial successes (progressive reputation cascade versus the result of a well orchestrated advertisement), immune system deficiencies (external viral/bacterial infections versus internal cascades of regulatory breakdowns), the aviation industry recession (9/11 versus structural endogenous problems), discoveries (serendipity versus the outcome of slow endogenous maturation processes), cognition and brain learning processes (role of external inputs versus internal self-organization and reinforcements) and recovery after wars (internally generated (civil wars) versus imported from the outside) and so on. In economics, endogeneity versus exogeneity has been hotly debated for decades. A prominent example is the theory of Schumpeter on the importance of technological discontinuities in economic history. Schumpeter (1942) argued that "evolution is lopsided, discontinuous, disharmonious by nature... studded with violent outbursts and catastrophes... more like a series of explosions than a gentle, though incessant, transformation."

### 4.3.2 Political and economic transitions

Consider the fall of the Berlin wall in October 1990 associated with a series of radical political changes in the Eastern Bloc. Over the period from 1989 to 1992, many east european countries engaged in a transition from a centrally planned economy to a democratic and market economy. Using agent-based model simulations and economic data, Yaari et al. (2008) discovered that all countries' GDP (gross domestic product) as well as other indicators of economic development (such as the number of privately owned enterprises) evolved through a generic J-curve, corresponding to a first phase of strong decay followed by a recovery and, for some countries, a transition to a growth rate surpassing significantly the levels under socialism before 1990. The first decay arch of the J-curve corresponds to the progressive demise of the "old centrally planned economy", whose shrinkage dominates the rise of the "new" free market economy (Novak et al., 2000). The second rising arch of the J-curve embodies the progressive transition to the "new economy" that burgeons as a response to novel conditions (Challet et al., 2009). In the case of Poland, Yaari et al. (2008) found that the new economy principally developed around a few singular "growth centers" associated with pre-existing higher education poles, which was followed by a diffusion process to the rest of the country. The centers of education were thus the main engines of the resilience and adaptability of the Polish nation to the new conditions. In contrast, other east european nations, such as Ukraine or even Russia, have fared much less well (Guriev and Zhuravskaya, 2009): for them, the transition resulted in a long lasting economic crisis that only recently has started to show observable improvement.

Let us scrutinize the economic transition in Russia. For a decade since the Berlin wall event, Russian GDP has been declining, with continuing huge drops in output and high levels of inflation. Russia went through a Great Depression more severe than that in the U.S. in the 1930s, with a decline in industrial production of over 60% from 1992 to 1998 (vs. some 35% decline in the U.S. Great Depression from 1929 to 1933), leading among many woes to the destruction of agriculture, deteriorating social conditions, health, education, environment, law, science and technology, high inflation and the destruction of the middle-class which is often the guardian of, as well as condition for, a functioning democracy. The Russian economy has been characterized over this time period as being riddled with crime and corruption. The transition was not to a market economy but rather to a criminalized economy, where the criminals established their own institutions in a process of self-organization (Intrilligator, 1998). The reasons for these problems have been identified (Intriligator, 1997; 1998): by endorsing a stabilization program of the russian economy based on liberalization of prices and the privatization of enterprises, the Yeltsin administration neglected the well-known but often forgotten fact that free markets requires strong institutions, and in particular a legal system, courts, lawyers, law enforcement; property rights, and so on, so that business contracts are enforced rather than subjected to the whim of the strongest. Moreover, a strong government is at the core of market economies, as shown by numerous anthropological and historical studies documented for instance in (Graeber, 2011). Russia's transition illustrates that externally imposed conditions, fundamental internal situations as well as a badly chosen design of governance (without institutions and working legal system) led to a new regime that has struggled for a very long time to recover and establish a functional state for the well-being of the people (Guriev and Zhuravskaya, 2009).

The so-called Arab spring that began in Dec. 2010 constitutes another telling illustration of our thesis. This revolutionary wave of demonstrations and protests occurred in the Arab world, leading to the ousting of the leaders of Tunisia, Egypt, Libya and Yemen and civil uprising in other neighboring countries. While media reports and scholars have often viewed the Arab Spring movements as positive steps towards more democratic governance, some skepticism is in order when examining the post-Gaddafi outcome in Libya for instance. Research at the NECSI ([necsi.edu)](necsi.edu)) suggests persuasively that the triggering factor for many if not most of the upheaval movements observed in arab as well as other poor countries around the world coincide with rapid and large rises of food prices (Lagi et al., 2011; Bertrand et al, 2012). Indeed, commodity prices more than doubled in 2008 due to a combination of environmental factors, the accelerating needs of booming countries such as China as well as speculation (Sornette et al., 2009a). As a consequence, world food prices skyrocketed, making many households' subsistence reach a crisis level. The inability of the governments of the concerned countries to cope with these stressors led to the transitions (or in many other cases to the search for the resolution of quite unstable states) to what can still be seen as evolving situations in search of an equilibrium. Whether the outcomes in Libya or Egypt are positive remains to be determined as the region has become very unstable and the future remains highly stressful and uncertain for most of the population.

This is reminiscent of the french revolution of 1789: more than enlightenment ideals, economic factors arguably played indeed a crucial role. As a result of bad harvest over most of the decade preceding 1789, a large part of the french population was exposed to strongly rising bread prices (the main food), leading to hunger and malnutrition. In the absence of adequate reactions by the government to the climate stresses that were adding to a very large national debt and an antiquated tax system weighting unfairly on the working class, the resulting discontent population became prone to push for major changes that culminated with the storming of the Bastille. Similarly to the situations resulting from the Arab spring, one should be cautious to claim that the extraordinary changes resulting from the food price stressors (among others) have always and systematically been for the better in all dimensions. The situation is perhaps best captured by the apocryphal statement of Chinese premier Zhou Enlai during President Richard Nixon's visit to China in February 1972: "too early to say" when referring to the assessment of the implications of the French revolution (he was in fact probably referring to the turmoil in France in May 1968 (Campbell, 2011)). Notice also that there is clear evidence that the French revolution has led to much bloodier wars in which whole nations have become involved in large scale conflicts involving many casualties (Cederman et al., 2011), showing again the relativity of the values of the regime shifts and their often unintended consequences.

These examples have illustrated two main points:

(i) the ubiquity of (rare) regime shifts due to the combination of abnormally large external circumstances (that are bound to occur in any nonlinear system if one waits long enough) and internal facilitating processes limiting the build-up of adequate resilience;

(ii) the value (in terms of economic consequences, change of well-being, moral level, culture) of regime shifts is open to debate, depends on the time horizon (beneficial short-term but detrimental long-term, or vice-versa) and is arguably relative;

All the examples treated in this subsection refer to situations in which scholars and observers would rate the pre-existing regimes as (to various degrees) undemocratic, oppressive and in opposition with the enlightenment ideals. As we shall elaborate in section 5 on recipes for resilience, much of the strength of a nation rests on the cohesion between its citizens that is called upon at times of stresses. In this respect, arab countries, the countries of the Soviet block, and France under the Bourbon dynasty developed modes of governance that embodied the roots of their demise, such as increased inequity and rigidities. One should not develop however the impression that this situation is a unique attribute of countries that do not embrace the modern western version of market economy and of democracy (which, by the way, is not a unique governance process of course but comes in many kinds and degrees).

Consider the situation of the largest western economies, including the United States of America, Japan and western Europe, whose indebtedness have reached, according to many analysts and pundits, unsustainable levels (Reinhardt and Rogoff, 2011). Scenarios for the next decades encompass the possibility for global critical transitions at worst or, at least, the need for massive readjustment of expectations (which is a polite way to say that retirees will get much less and after working significantly longer, average social coverage will shrink much further, standard of livings will at best plateau with many signs of deterioration for the median household). Here again, one can argue that the western economic systems have been build on a model of run-away indebtedness that, on the "short term" of the past several decades, brought extraordinary gains, at the cost of increasing systemic and global risks (Sornette and Woodard, 2010). The on-going crisis of debt-strangled european nations is far from finished, as nothing has been done in depth to address the problems of insufficient growth of productivity and innovations (Sornette, 2010), of the demographic bottleneck, and of reigning on wasteful over-spending beyond one's means by addicted consumers as well as nations spoiled by the failure of democracy replaced by demagogic politics (Gore, 2007). The US should not be forgotten either, if only because its financial system is effectively bankrupt, but held artificially alive by rounds of buying toxic assets by the Federal Reserve and the successive spells of so-called quantitative easing. An even greater crisis if possible is probably awaiting Japan, which relies on the policy of essentially zero-interest rate in order to cope with a total debt that dwarfs that of all other nations. The policy of ultra-low interest rate seems to become the new reference point of debt-strangled nations in order to be able to honor their interest payments, which yet not fully appreciated consequences concerning the transfer of wealth between generations and the possibility to face the huge retirement liabilities. Globally, the diagnosis is clear: these systems have built economic organizations that contain in themselves the seeds for monstrous systemic instabilities towards major re-organizations. The 2008 US crisis and the 2010-2012 sovereign european debt crisis are probably nothing but the premises of much more significant crises at the global scale. Such a prediction is warranted on the observation that none of the real causes of the crises have been addressed and only superficial short-term remedies have been offered until now (Mauldin and Tepper, 2011; see also chapter 10 of Sornette (2003) which is based on Johansen and Sornette (2001) and, more recently Akaev et al. (2012)).

Thus, we can add to the two points (i) and (ii) above a third one:

(iii) social and political systems seem to be intrinsically unstable on the long term, building up internally the mechanisms of increasing vulnerabilities via the very processes that seem initially the most favorable.

Resilience is therefore a fundamental question that needs to take into account both the conflicts between time scales (generations) and the unintended consequences of short-term innovations and improvements (Ferguson, 2011).

**4.4 Debunking "anti-fragility"**

It is appropriate to end the present section, discussing whether stress can be beneficial, by the extreme view proposed by N. Taleb (2012) summarized under the vocable "antifragility". According to this concept, "antifragile" systems may not only resist and recover efficiently from stressful events but may actually benefit from them in very direct ways and on the short term. Taleb lists a number of examples illustrating this view: muscles and bones, owning insurance or financial derivatives, decentralized organization and so on. If correct, the antifragile concept would contradict our whole construction presented above. To understand the source of disagreement, we now dissect Taleb' s proposition. In a nutshell, antifragility describes the quality of some systems that are designed to profit from particular stressors that produce stress in other systems and to which they are not sensitive themselves. But, as we are going to show, these so-called "antifragile" systems have also their own vulnerability to other stressors that lie outside their tailored design.

### 4.4.1 The put option paradigm

The example that captures the essence of the whole "antifragility" argument is that of financial derivatives. Consider specifically a put (also called "sell") option written on some underlying financial asset. The later has values that fluctuates more or less randomly, with sometimes large excursions in the positive (gains) as well as negative (losses) ranges. An investor owning this asset will be exposed to possible rare large losses, the so-called tail events. The investor's investment is thus a priori vulnerable to the occurrence of financial shocks that may hit his asset and make it fall abruptly. Fragility is particularly acute if, as such time, the investor needs to cash out for some consumption needs (unforeseen medical expenses or student university tuition for his children) at the much lower asset value following the crash. Another investor, who has bought a put option of that same asset, has a diametrically opposed perception of the situation: when the asset plunges, the value of his put option sky-rockets upwards. In the terminology of antifragility, the put option investment of the second investor is antifragile, since it profits from large negative price movements that hurt most other investors. The put option paradigm is actually underpinning the whole antifragility concept when applied to general situations, as developed in (Taleb and Douady, 2012). To summarize, Taleb advocates strategies and policies that construct effectively put options everywhere!

Let us clarify how a put option works. First, it needs a risky asset or a basket of risky assets that are subjected to the influence of many natural and social factors so that its value fluctuates with sometimes large amplitudes. Second, it needs a counter party, say a bank, which accepts to create the put option and sell it to the second investor. In the case when the put option is exercised, the counter party has to pay for the gain of the option owner. The put option strategy is thus conditional on others taking the other side of the risks.

It is important to realize that the put option strategy is built on the premise that it can only work when endorsed by a minority of investors, at the expanses of the others. Take the example of the so-called "portfolio insurance" strategy developed in the 1980s by Leland and Rubinstein. Large institutional investors wanted to insure their large portfolios against possible drops of the stock market. For this, the simplest and most efficient strategy consists in buying put options on the assets held in the portfolios. However, the sheer volume of put options needed was beyond what banks and other option writers would be able or willing to offer. Or, if offered, the requested prices would have been prohibitive. Leland and Rubinstein then used the replicating construction of the Black and Scholes option pricing formula to devise a simple and effective way of constructing synthetic put options just based on the underlying assets and on bonds. The synthetic put options thus created led to a flourishing business where, at the time just before the crash of October 1987, more than one third of all US institutional investors had implemented the Leland-Rubinstein so-called insurance portfolio strategy (MacKenzie, 2008). The weakness of this whole construction however was revealed as markets started to stumble the week before "black monday" 19 October 1987. Because the synthetic put options operate by selling the underlying stocks when the later decreases in value, as the stock values start to go down, the synthetic put option strategy led to sells, pushing prices further down, these losses aggravating the negative sentiments of the markets, leading to an avalanche of sells reinforced by the technical implementation of the synthetic put options leading to a vicious positive feedback to the bottom. After the crash of October 1987, many pundits and scholars have concluded that, with a large probability, synthetic put option strategies were responsible for aggravating strongly the severity of the crash (Barro et al., 1989). What was supposed to be a bullet-proof strategy turned out as a catastrophe due to its hidden vulnerability with respect to synchronization. In other words, buying put options works when you are in the minority and no collective herding behavior occurs. More generally, the whole business of insurance is based on diversification of exposures. This message was vividly brought home to major insurance and re-insurance companies in the aftermath of the 911, when the capital stored in stock markets needed to be sold to compensate clients for their losses plummeted at the same time. This illustrated another mechanism of fragility of the supposed antifragile insurance strategy.

The 15 September 2008 Lehman Brothers bankruptcy and 16 September 2008 AIG official bail out demonstrate another fundamental fragility of the antifragile put option strategy. In short, major investment banks around the world had invested in CDO (collateralized debt obligations), which are securitizations of mortgages offered to millions of american households. Many of these investment institutions search for ways to insure their exposition to possible losses on the CDOs by buying massive amounts of CDS (credit default swaps) from counterparts, the most famous and by far largest being AIG, the then largest insurance company in the World. Different from what their name suggests, CDS work essentially as put options paying large amounts when the underlying CDO loses value and/or when some trenches of the CDOs start to default. Buying CDS was a perfect antifragile strategy to profit from the rather visible problems looming as a result of the enormous real-estate bubble that has developed in the US from the early 2000s to 2007. Except for one thing: the credit risk of AIG was not considered. Default of AIG was inconceivable. The problem is that the collective use of the antifragile CDS strategy led to such an enormous exposition of AIG to a downturn of the US real estate market that its total capital base became insufficient, finally leading to its quasi-bankruptcy and its final salvation by a massive injection of capital from the US treasury and a consortium of investment banks. The so-called antifragile CDS strategy backfired to systemic proportions, whose real consequences are still to be solved as the time of writing.

Moreover, for an inner circle of investment banks, the CDS strategy turned out to be really profitable, though not from the intrinsic structure of the strategy but from playing the fear to the public of a global financial and economic meltdown as well as from using high-level political connections. The bail out packages, which were put in place in September 2008 and following months, ensured the payments of most of the liabilities at 100% face value (which AIG could not longer support) to the major investment banks. The weight of these payments was in the end supported by the taxpayers.

In sum, these dramatic examples illustrate that antifragility does not exist. In general, for systems subjected to variability, noise, shocks and other random perturbations, it is possible to develop strategies that, on average, benefit from variability, but not any variability. Such strategies are designed to profit from the variability of particular stressors. Simultaneously, they are vulnerable to other stressors. The refusal to accept this fundamental characteristic (or intrinsic weakness) shared by any strategy or system is very dangerous, as it may lead to unexpected shocks or intended manipulations by insiders. For instance, in the financial sphere, antifragility is a name for the exploitation of a situation that turns losses for most into gains for some by special design, which is however vulnerable to non-anticipated occurrences. Moreover, the so-called antifragile strategy can contain the germs for large externalities, leading to systemic crises for which neither the strategy itself nor the system are prepared for.

### 4.4.2 Can antifragility be beneficial itself?

Taleb has provided many tentative examples of supposedly antifragile systems, putting them in contrast with fragile and robust systems. For each instance (i-vii) below, the antifragile system (according to Taleb) is indicated in boldface and contrasted with its opposite fragile version:

(i) civilization (**nomadic and hunter-gatherer tribes** versus post-agriculture modern urbanization);

(ii) production (**artisans** versus industry),

(iii) science/technology research (**stochastic tinkering** versus directed research);

(iv) nature of the political systems (**decentralized political systems** versus centralized nation-states);

(v) decision making (**convex heuristics** versus model-based probabilistic approach);

(vi) literature (**oral tradition** versus books and e-readers);

(vii) reputation (**artists or writers** versus academics, executives and politicians) and so on.

In all these examples, one notices that the antifragile system is much less productive that its fragile counterpart. In example (i), the capacity to support larger and growing populations has received an enormous boost with the introduction of agriculture while hunter-gatherer tribes had zero or very small growth. A typical north american family now commands a quantity of artifacts on par with or larger than that of a pharaoh at the peak of the classical pharaonic civilization. This illustrates that, in example (ii), the elaborate supply chains of modern industry based on the collaboration between millions of workers delivers enormously more than the whole summed contribution of individualistic generalists. In example (iii), the classical Greek tradition let place after many centuries of "stochastic tinkering" to an organized scientific production in the last few decades that dwarfs absolutely the knowledge accumulated earlier. In example (iv), nation-states have been able to mobilize resources unheard of decentralized political systems. Clausewitz (1984) [1832] in his classic book "On war" observed that the french revolution introduced the nation state, which led to global wars with enormously more resources, an hypothesis recently supported quantitatively using statistical comparative history (Cerderman et al., 2011). In example (v), heuristics may often work for simple everyday problems and when immediate quick-and-dirty solutions are required, but would be unreasonable for decision making and management in sophistical modern systems dealt with by surgeons, airline pilots or technicians of nuclear plants. In the case of literature (vi), it is clear that oral tradition would not fail if electricity is no more available but, on the other hand, it is a very inefficient and low-density information medium, quite unsuitable to share and store the explosive amount of modern knowledge. Lastly (vii), academics, executives or politicians have developed extraordinary specialized skills that are (in principle) translated into positive reputation. A positive reputation serves the goal of producing more or delivering higher quality services and/or of being trusted. In contrast, some artists and writers just need any type of reputation as long as people and media speak about them, because their business is in a sense to bank on their fame.

Pushing Taleb's reasoning to the extreme, one could conclude that being a beggar is one of the most desirable antifragile state to be in, since the person has nothing to lose and can only benefit (if he survives) from any change of his position. The condition "if he survives" actually demonstrates the essential hidden assumption underlying antifragile examples. Otherwise, as soon as there is something to loose, to disproof or the possibility of a disfunction, as when owning assets, possessing a reputation, using a decision model, or production scheme, there are many additional stressors that could cripple the system. Being rich, young, healthy, beautiful and loved is the ultimate fragile state, but who would exchange it for its absolute antifragile poor, aging, ill, ugly and lonely alter ego.

### 4.5 Can stress be beneficial? Our answer

To summarize, we have shown that stress is unavoidable and that systems co-evolve with their stressors. The survival of a system depends on its ability to cope with and adapt to numerous stressors. In this sense, the life-span of the adapting system is relatively longer than those of many of its stressors. These stressors, coming one after another, are progressively shaping the system, demonstrating sometimes a true symbiosis and an astonishing emergence of new features that can be beneficial for the system itself. In evolutionary biology, non-visible or "neutral" mutations occurring in the presence of internal stresses as well as small external stochastic perturbations, and which leave fitness unchanged, are considered beneficial because they improve the system's robustness (Kimura, 1983; Ciliberti et al., 2007). They provide a diversification by enlarging the toolbox of defense without disruption and prepare for major jumps when necessary or when ready (Wagner, 2005; Ciliberti et al., 2007). This concept seems to have broader applications, as recently proposed to quantify software robustness (Schulte et al., 2012). Finally, extreme stressors are relatively rare events, but they play an exceptional role in creating the global landscape and activating the mechanism of natural selection. Their magnificent power gave rise to legendary names - "dragon-kings" (Sornette, 2009; Sornette and Ouillon, 2012), for the extreme stressors of endogenous nature, and black swans (Taleb, 2007) that are characterized by exogenous sources.

The response of a system to stressors depends on the level of stress within it. To make the system more efficient and flexible, it is important to learn how to use normal stress as a signal of on-going changes and as a guide for needed adaptation to better fit to the evolving conditions, so that a system can survive and benefit from them. In the presence of extreme stress, resilience, that is, conservation of the status quo, may not be anymore an option and the resources should be directed towards an unavoidable transition to a new regime that can bear or even profit from the stress: in the words of Giuseppe Tomasi di Lamedusa, in The Leopard: "If we want everything to stay as it is, everything will have to change."

In Section 5, we propose strategic principles for system resilience and describe some of them in details. However, the adoption of strategic principles in most cases would require global systemic changes and would face numerous difficulties, partially described at the end of section 3.3. Therefore, in Section 6, we discuss some of these limitations and propose original operational solutions.

### 5-RECIPES FOR RESILIENCE

### 5.1 Generic recipes for resilience

The systems that were previously mentioned are very different, and so are the conditions of their functioning and the stressors they face. Nevertheless, from the fact that stress is a non-specific response of a system that depend weakly on the type of stressor, it derives that the development of generic recipes to cope with stressors is both possible and crucial for strengthening its resilience.

We propose the following brief synthesis of strategic principles for the sustainable development of any system, which borrows from a variety of risk management thinkers, from Sun Tzu's "The art of war" (circa 500 BCE), Clausewitz' "On war" (1984) [1832], John Boyd's "certain to win" strategy and his OODA (observe-orient-decide-act) loop (Boyd, 1986; Richards, 2004) and Sheffi (2005). While rooted in ancient wisdom, their modern framing and phrasing do not diminish their reach and eternal relevance.

1) Develop strategic vision; orientation and focus on the present and future, and not on the past; establish **clear goals** (subsection 5.2),

2) build up, through investment and/or education, **fundamental values, right incentives and fair remuneration** (subsection 5.3),

3) diversify and promote **heterogeneity**, as well as **decoupling** of key components for sufficient redundancy,

4) develop operational mechanisms to **enforce contracts**,

5) promote **transparency, communication and ethics**.

At the operational level, tools for quantification of stress signals and learning from them should be put in practice in order to cope with stress effectively, i.e. to improve (i) the quality of decisions in the presence of risks and (ii) the management of resilience. These tools are to serve the following goals:
        a) development of individual strengths together with awareness of one's limits,
        b) promotion of collective action and collaborations,
        c) analysis and classification of stressors,
        d) risk identification and tracking,
        e) continuous measurements and diagnosis of endogenous instabilities,
        f) never ending verification and validation,
        g) always keeping on edge by questioning assumptions and existing processes.
This last point is easy to formulate on paper but much harder to implement in practice, if only because of the common adage that "No one see any pressing need to ask hard questions about the source of profits, of success, or stability, when things are doing well." Building resilience requires indeed a kind of paranoic obsession that things could go wrong, when everything appears to be fine. Sections 5.3.3 and 5.4 provide concrete examples of such operational tools.

**5.2 Formulation of goals and objectives**

The first step on the way towards implementing the strategic principles for the sustainable development of a system is to identify and spell out the goals and objectives, which can also be called utility functions of the system. In this subsection, the strategies and methods of resilience growth are outlined into accordance with different types of goals.

(1) At the most basic level, a first goal is to ensure survival, which calls for the measures promoting viability that are described in section 3.2, in particular using stress as information and being always ready for managerial actions to ensure that the system remains in its basin of attraction.

(2) A second type of goals is often the conservation of the status quo, of existing wealth, of present standard of living. This triggers what we referred to as the "fight" response, which applies when the stress is significantly smaller than the existing resilience of the system. However, many systems, human societies and organizations in particular, reach high levels of wealth, which were obtained at the cost of strong optimization, decrease of reserves, indebtedness, increase of inter-dependencies (Diamond, 2004), which result in loss of resilience. In these situations, the fight response to maintain homeostasis at such high development levels is simply not possible in the middle and long term, because even small stressors will in the end be enough to trigger a change of regime due to the endogenous build-up of a critical fragility. As a vivid and painful example, one can argue that the present on-going sovereign debt European crisis belongs to this class. Only with a profound reassessment of goals taking into account the realities of the globalized economy and the structural unbalances underlying the artificial construction of the euro dream, can one hope to address the systemic nature of the European conundrum.

(3) A third type of goals, often observed in high-tech industries for instance, is for an entity to become and stay the leader among its pairs, hence developing highly competitive attitudes and strategies. IBM, Toyota and Apple are different examples of firms that were able to get to the top and remain there for longer than thought initially possible. For IBM, this was through its evolution from a mainframe computer hardware company to a service provider offering all possible integrated solutions to a large range of customers, thus redefining continuously what is the essence of being IBM. For Toyota, the empowerment of the factory workers, instructed to focus on the delivery of just-in-time products, led to a remarkably motivated and productive workforce delivering high quality products for more than 50 years. But the 2010 car recalls due to the sticking accelerator pedals and failing electronic throttle controls demonstrated that bureaucracy, overconfidence and weak management have lately underpinned Toyota's fall from grace.

Apple's remarkable success can be attributed to its focus on innovation aimed at surprising and enthusing customers, by functioning as a secret organization with a self-perpetuating start-up culture. For these companies, resilience at the top requires internal engineering of their ever on-going mutation, aiming at shaping the future rather than reacting to it, in the spirit of "You don't wait for the future. You create it." (Hwang Chang Gyu, 2004).

(4) In the modern world, the economic language and agenda dominates, with such concepts as utility function (assumed to capture people's goals) and growth of GDP (gross domestic product) taken as the universal measure of improvement and success. But, too little attention is given on what the US founders enshrined in the US constitution as one of the three main goals of well-functioning societies, namely the pursuit of happiness. In the United States and in many other industrialized countries, happiness is often equated with money. This simplifying assumption provides a convenient way of quantifying and comparing heterogeneous preferences of different agents within a unifying framework. This money (or economic utility function) approach has shaped our culture. Only the small Himalayan kingdom of Bhutan has made its priority to grow, not its GDP, but its GNH (gross national happiness). According to King Jigme Singye Wangchuck, Bhutan's goals are to ensure that prosperity is shared across society and that it is balanced against preserving cultural traditions, protecting the environment and maintaining a responsive government. In our context, this can also be interpreted as promoting a resilient society, based on (i) robustness anchored at the individual level (a happy and balanced person is arguably more robust in her behavioral response to stressors) and (ii) through cohesion within the society build on a common understanding that ethical behavior is fairly rewarded and equity (and not "equality" as in communism) is the standard reference.

The development of a strategy requires an out-of-the-box thinking and the consideration of multi-dimensional objectives. Setting up goals depend also often crucially on the time scales of interest as well as on the size scales (individual versus group versus society). There are well-known differences in goals and welfare attained at the individual versus collective levels. It is often difficult to reconcile the preference of individuals with those of the aggregate group. This is known as Arrow's impossibility theorem in social choice theory (Campbell and Kelly, 2002). At the extreme, the sacrifice of individuals may ensure the survival of the whole system. Lymphocytes are not resilient individually but ensure the resilience of the immune system. Such strategies are apparently at the opposite end of Bhutan's emphasis on individual happiness. This suggests that there may be several paths towards system resilience and/or that the level and type of resilience is also a matter of choice, given the conflicting requirements (costs versus benefits at different levels).

**5.3 Fundamental values and individual strength as a basis of resilient societies**

Resilience of a system depends first on the sustainability of its development and its fitness to environmental conditions through processes of adaptation and co-evolution (as described in section 4.1 and 4.2). These characteristics ensure reduced amount of stress accumulating within the system. Second, the resilience of the system is based on its strength to cope with the existing stress and its causal factors. To grow and maintain both of these characteristics over the long term, it is essential to build up fundamental values through education and investment, and to implement the right incentives and fair renumeration.

At the system level, it can be illustrated by the following examples:
- fundamental prices of assets are more stable and predictable than their bubble components, which are unstable and may lead to severe crashes;
- practical skills (farming, engineering, programming, the development of the real economy, and so on) should be better rewarded both economically and in our cultures; stakeholders should pay attention to the added-value of supporting services (financing, marketing, management, and so on) and not hesitate to shrinking and redirecting efforts when these supporting services become tyrants rather than servants of the real economy;
- hard work, persistence, tenacity and dedication should be emphasized (which is at the opposite of the common modern emphasis on the role of chance and luck, the belief in easy profits, the "american dream" now fueled by a perpetual expanding credit engine).

The implementation of the recipes for resilience designed at the system level may not all apply directly to the individual, due to differences in the goals as well as psychological and physical aspects. The rest of this subsection is focused on recipes for personal resilience and top performance, which are easy to implement by everyone. To change the world, one should start with oneself.

Section 3.3 documented that many natural systems evolved to function "at the edge of chaos", characterized by a sharp balance between level of risks they face and costly resilience build-up. Management of social-economic systems is also striving to achieve a balance between costs of increased resilience and its benefits. But would "at the edge of chaos" be a desirable state for a human? To stay a long time close to criticality, in a kind of alarmed position, requires constant attention, give rise to worries and triggers anxiety.

In the end, there is the possibility that such a critical state does not lead to an efficient allocation of resources of the body and mind, but becomes stress itself.

One should consider an additional dimension, an often neglected benefit that comes from higher resilience: resilient people are more "happy" and vice-versa. Indeed, people who feel on top of their life and who can face stress are more relaxed, enjoy more the present and live longer. More resilience promotes a more positive attitude to one's own life and to others. In contrast, those of us who are in a continuous race to face the constraints of personal and professional life live in a state of anxiety, a condition that has been accelerating in severity in recent decades as witnessed by the exploding sales of antidepressants. Research in psychology and psychiatry confirms the existence of a strong interdependence between resilience and happiness, with positive feedback loops in which higher positive mind set promotes resilience and vice-versa (Jackson and Watkin, 2004; Srivastava and Sinha, 2005; Cohn et al., 2009). In particular, positive emotions help people build lasting resources (Cohn et al., 2009). And it is how we respond to stress and hard time that determine our successes or failures, rather than the nature of the stresses themselves. This supports again the need for generic and robust recipes for building up resilience and... happiness at the individual level.

In a review covering a large body of research investigations on individual resilience, Coutu (2002) extracted the three main characteristics that are most often associated with resilient people:
  (i) a staunch acceptance of reality,
  (ii) a deep belief that life is meaningful, and
  (iii) an uncanny ability to improvise.
Our own experience and reflection suggest to add
  (iv) the ability to keep an inquiring mind that questions assumptions and the status quo and
  (v) a strong belief that our project and endeavors will succeed.
The seven factors of resilience reviewed by Jackson and Watkin (2004) from the psychological point of view overlap with the two first items, that is, with the need of developing a realistic view of reality and finding meanings (or causality). Indeed, they cite the following seven factors: (a) emotion regulation, (b) impulse control, (c) causal analysis, (d) self-efficacy, (e) realistic optimism, (f) empathy and (g) reaching out.

These are descriptors or traits of resilient individuals. In order to be genuinely useful however, the next step is to identify whether and how it is possible to acquire, nurture and augment these traits. We are here entering the controversial domains of psychological programs and even psychiatric treatments. We take a simple "mechanistic" approach based on the premise that the above traits do not reside in a vacuum but rather are properties of bodies and minds that can be trained. Take the example of will power. In a study of one million people quoted by Baumeister and Tierney (2012), most said that self-control was their biggest weakness. So can people build up their willpower? Or are some people just born that way? In their recent book, Baumeister (who directs the social psychology program at Florida State University) and Tierney (2012) argue that willpower is like a muscle, and like all muscles, can be exhausted through overuse, but also trained to be made stronger. We could say that a strong willpower gives benefits by a slow accumulation of small gains that grow over time. The build-up of willpower operated via a positive feedback process: the more you have, the more you use "rituals" and checklist type approaches, the better the performance, the stronger is gratification for the efforts spent, the larger the willpower, the more .... in a virtuous loop of self-reinforcement. Baumeister and Tierney also emphasize that everything is linked together and that one energy resource is used for all kinds of acts for self-control. One could then argue that, by training and augmenting the energy source, the stronger and more energetic the body and the mind, the easier it is to develop the factors promoting resilience. In this strategy, resilience has its underpinning in the strength as well as cohesion between constitutive elements found at the level of metabolism.

In a recent contribution, one of us (Sornette, 2011) has laid out seven governing principles for personal resilience and performance that we repeat for completeness. We refer to the original essay and its detailed documentation and argumentation. The seven guiding recipes for individual resilience and performance are anchored in processes that control our biological and psychological well-being. Implementing these principles require willpower, which can be augmented both by the fact of being used, as in the muscle analogy of Baumeister and Tierney (2012), and by promoting the access to more energy as the source for action.

1) **Sleep**: Rest with quality sleep for a minimum of 7-8 hours per night;

2) **Love and sex**: Cultivate the romance and relationship with your special partner; interrupt your work when needed with one minute of intense focus on the loved one, perhaps using romantic pictures of him/her to trigger happiness hormones that boosts brain performance and well-being.

3) **Deep breathing and daily exercises**: Start each of your day (no exception) with 5-10 minutes of exercises, including deep breathing-stretching followed by abdominal and finishing with a very short intense workout; perform a few 2-3 minutes of intense workouts and deep breathing at different times of your day in your office or wherever you happen to be in order to oxygen your body and refresh your brain;

4) **Water and chewing**: Drink at least 2 liters of water per day (no canned juice, no coke, no beer, no sugar) outside meals and drink minimally or not at all during meals (a small glass of red wine or cup of hot green tea is fine); "drink your food" and "eat your drinks".

5) **Fruits, unrefined products, food combination, vitamin D and sun exposure and no meat and no dairy**: Eat as much fruits with water as possible on an empty stomach during the day, avoid meat and consume only unrefined products and cereals; avoid bad food combination to avoid conflicts between alkaline versus acid foods.

6) **Power foods**: onion, garlic, lemon, kiwis, almonds, nuts, dry fruits for super-performance in time of intense demand.

7) **Play, intrinsic motivation, positive psychology and will**: rediscover the homo ludens in yourself in things small and large so that work and life become a large playground, cultivate motivation as a self-reinforcing positive feedback virtuous circle.

## 6-HUMAN LIMITS AND OPERATIONAL SOLUTIONS

### 6.1 Intrinsic human limits

#### 6.1.1 Identification of stress signals and reactions to them

The analysis of the major industrial catastrophes, such as the 1986 Challenger space shuttle disaster, the explosion of the Ariane V rocket on its maiden flight in 1996, the Deepwater Horizon BP oil spill disaster that started on 20th April 2010, the Fukushima-Daichii nuclear accident in March 2011 and so on, reveals common problems in the following areas:
1) gathering information;
2) aggregating and communicating data;
3) maintaining a state of attention.
These same issues, which have been documented as underlying causes of these dramatic events, are similarly found underlying most accidents and crises in different fields of human activity, including the financial crises that started to rock the world in 2007-2008.

Gathering evidence about informative incidents is a well-known challenging task in the practice of operational risk management. Employees often experience a conflict of interests with respect to reporting problems concerning the area of their own responsibility or those of their colleagues. This may rise, for example, from the fear of punishment, disapproval of colleagues and seniors, and increase of duties to correct revealed weaknesses. As a result, signals of stress are often lost, near misses are not recorded, forgotten or dismissed, and decisions are made on the basis of unrealistically optimistic data. Furthermore, from the failure of reporting and aggregating information that is in fact known within the organization, vulnerabilities are accumulated and lead to greater accidents.

The other side of the "information problem" lies in the difficulty of maintaining a constant state of attention or excitation. It is not enough to detect a signal of growing stress, but there should be measures taken to address the issue. Unfortunately, people get used to warning signals and false alarms, and lower their guard. Again, this applies to all the above mentioned industrial catastrophes and to many more.

The first step in dealing with these problems is for the top-management to accept the unavoidable nature of stress so that appropriate stimulating mechanisms can be developed:
1) for gathering and communicating information:
- no punishment for self-reported occasional misses, as well as in the cases when all sufficient measures were taken to ensure a desired result (i.e. evaluating the process of decision-making, but not only an ex-post outcome);
- confidentiality;
2) for maintaining a high attention:
- "zero tolerance" to controllable misses;
- the introduction of random stressors (such as sending "fake hard customers" to check the professionalism of employees);
- a rewarding system for catching a stress signal.

### 6.1.2 The "logic of failure"

In their study on the "logic of failures" (Dörner et al., 1990; Dörner, 1997), Dörner and collaborators have found that there is indeed a logic in the origins and processes leading to failures, in the sense that (a) humans experience failure more often than success when intervening in complex systems, (b) the failures are not random, but exhibit common patterns and (c) the understanding of these patterns offer operational rules to prevent the failures. The studies performed by Dörner et al. led them to formulate general recommendations taking the exact counterpoints of the negative behaviors and habits that tend to inhabit people. Unsurprisingly, these recommendations overlap and sometimes complement the generic recipes outlined in section 5.1. In order to avoid failure and develop successful management of complex systems, one should

(a) continue to reflect and ask questions during the evolution of the project or system,
(b) act after careful analysis and be multi-faceted to ensure a rich toolbox of responses,
(c) strive to anticipate effects of one's actions,
(d) estimate possible negative feedbacks and unintended consequences,
(e) not shy away from adapting policies that are not working, and
(f) carefully assess the real goals as opposed to be over-involvement in pet projects.

### 6.1.3 The "illusion of control" syndrome

Last but not least, one should always have in mind the "illusion of control" syndrome (Langer, 1975; Satinover and Sornette, 2007; 2011), as already mentioned in the introduction. As a corollary, individuals appear hard-wired to over-attribute success to skill, and to underestimate the role of chance, when both are in fact present. Grandin and Johnson (2005) recount experiments pitting humans against rats, in which the humans, like the rats, have not been explained the rules of the game but must infer them from the situation. In such experiments, rats often beat humans, because humans tend to over-interpret randomness and find meaning in random patterns. Normal people have an "interpreter" in their left brain that takes all the random, contradictory details of whatever they are doing or remembering at the moment, and smoothes everything in one coherent story. If there are details that do not fit, they are edited out or revised for sense making, providing a powerful mechanism for the illusion of meaning and of control. These phenomena are ubiquitous. Langer (1975) summarized the problem in a rather amusing way: "normal people's high level of general intelligence makes them too smart for their own good."

This problem is perhaps best illustrated in finance where, after a full cycle of rise and fall after which stocks are valued just where they were at the start before the fall, most investors lose money by over-reacting and thus selling close to the bottom before the rebound (Guyon, 1965). More recently, a very large body of academic works support the conclusion that most managers underperform the "buy-and-hold" strategy and that the persistence of winners is very rare (Malkiel, 2012). Nevertheless, managed funds and the demand for professional investment advice has never been stronger and is a multi-trillion dollar industry, dominating the world of pension funds, mutual funds, sovereign funds, private banking and so on. [Disclaimer: we are not advocating buy-and-hold or inactive strategies, because we believe that the second decade of the 21st century is characterized by disequilibria and instabilities that make extrapolation of the past at best uncertain and mostly misleading.] The "illusion of control" syndrome is thus a call for realizing and understanding our cognitive biases (just look at http://en.wikipedia.org/wiki/List_of_cognitive_biases for an impressive list compiled on wikipedia). The psychological as well as philosophical literatures have discussed many times the intrinsic limits faced by any investigator trying to determine whether and how her own cognitive processes may deform her knowledge construction of the "outside" world. This is typified at the extreme by the madman who concludes, from the deformed lenses of his perceptions, that it is the rest of the world who is mad. In the context of dynamical game theory, Satinover and Sornette (2007; 2011) have determined precisely the conditions under which the "illusion of control" syndrome occurs. In dynamical first-entry games (a subset of game theory), they found that low entropy (more informative) strategies under-perform high entropy (random) strategies. This typically occurs in situations where there is a large amount of randomness, of uncertainty as well as the presence of negative feedbacks of the decision makers' actions onto the system.

### 6.2 "Crisis flight simulator" for management of complex systems and resilience build-up

The "illusion of control" and the "logic of failure" raise the following fundamental questions for practice. What is the value of management? How much management and control is needed? How can we falsify the value of control and of management, given that we do not have the luxury of playing history twice or multiple times? How is it possible to improve management skills when dealing with complex systems? Many studies and thinkers have pondered these issues. The recommendations given in the literature argue for a balance between extremes, such as strong top-down leadership to convey the goals and the vision, together with large responsibility and autonomy given to the bottom execution; a cohesive and strong backbone linking the individuals in an efficient hierarchical network of complementary abilities and trust together with a flexible adaptive organization to face changing and uncertain conditions. But how to achieve the right balance?

We propose that the answer lies in fostering a permeating and ubiquitous learning and testing environment, as occurs during academic curricula, and which should grow within all resilient organizations. This can take the shape of the systematic development of "crisis flight simulators" everywhere.

Consider the subprime crisis that started in 2007 with epicenter in the U.S. and the on-going sovereign crisis in Europe. To stop these systemic crises, central banks and governments have resorted to extraordinary measures, such as growing the balance of central banks with amounts of so-called toxic assets at levels dwarfing all known historical precedents. It is fair to state that we now live in a world where central banks and government are performing experiments in real time that are impacting billions of people, based on dated economic models (such as the Dynamical Stochastic General Equilibrium), which until recently did not even incorporate a banking sector and could not consider the possibility of systemic financial failures due to contagion. Not much has changed, though. The "primitive" approach of policy and decision making, based on rule-of-thumbs, political agenda, demagogy, and untested models, is still in full force. In contrast, we argue that progress requires to endow decision makers with tools to learn and to practice at the level that airline pilots or surgeons already experience in their training. These "flight" or "surgery" simulators reproduce as faithfully as possible real processes as well as all imaginable and even unimaginable scenarios to perform "what if" exercises. This approach is relevant for all kind of decision markers, including those in the financial, policy, engineering and environmental domains, and concerns also the public, students and anyone interested and responsible.

A good example of an early development of "crisis flight simulators" is the approach of Dörner et al. (1990) and Dörner (1997) mentioned above. Dörner and his colleagues conducted experiments with computer simulated environments, which included two groups of participants - executives and students. Analyzing the results of the experiments and the significant better performance of the executives, the authors proposed the concept of "strategic flexibility", which is essential in coping with uncertainty and can be learnt through practical experience or by successive computer simulations.

The goal should thus consist in developing sophisticated convivial simulation platforms that incorporate detailed physical, geological, meteorological, geological, architectural, sociological, cultural, psychological and economic data with all known (and to be tested) feedback loops. For a given simulation, decision makers are given the power to make decisions on allocated resources to develop projects and to mitigate risks according to different strategies. The simulations will then demonstrate the consequences of the decisions within a multi-period set-up. Only by "living" through scenarios and experiencing them, can decision makers make progress. For instance, there is enormous evidence in the laboratory and in real life settings that veterans who have lived through financial bubbles and crashes, through environmental crises and so on, are much better at prevention and mitigation. But, in practice, the cost is too large to learn from real life crises. This calls for a methodology for resilience based on the development of simulators that decision makers use to understand the complex dynamics of out-of-equilibrium systems whose behavior intrinsically includes changes of regimes, bifurcations, tipping points and their associated crises. This ambition is for instance shared by the FuturICT project, as embodied in its "Living Earth Simulator", which aims at enabling the exploration of future scenarios by large-scale simulations and hybrid modeling approaches running on supercomputers [Bishop et al., 2011; Helbing and Balietti, 2011; Helbing et al., 2011].

With such tools, the decision maker is able to understand holistically the dynamics of the system, in a systemic way, which means that he can understand the existence of systemic instabilities as one of the dynamical solutions of the system evolution. This must be complemented by a classification of the different regimes possible, a phase diagram in which the decision maker understands which control leads to the region of the unwanted regimes and which do not. He needs to understand that bifurcations and changes of regime are a natural and expected part of natural and social systems. This understanding does not occur via studying arcane mathematical theory but, instead, by experimenting as in real life, albeit with the protective comfort of the simulator and the efficiency of scaling space and time as needed. Only under this systemic structural understanding, can he interpret correctly the precursory signs in real life and use them to correct and steer the system towards resilience and sustainability.

In order to achieve effective "crisis flight simulator" platform for management and resilience, three technical goals must be achieved: (i) modeling, (ii) collective action and (iii) crowd sourcing.
    First, there is the need to transform complex risks scenarios from natural language into a logical, machine-interpretable description. For that, it is necessary to reach a sufficient level of abstraction to address a broad variety of scenarios and make them reusable. We envision that complex risk scenarios could be seen as electronic circuits with components acting as relays, delayers, amplifiers, dampers, transistors, and so on, connecting at-risk entities. For instance, consider three entities A, B and C. A transistor dependence would be: A fails implies that C fails if B is activated. By combining basic components, arbitrarily complicated scenarios can be built and, moreover, scenarios can be machine-tested. This first

approach intends to identify elementary components from which any arbitrarily complicated risk situation can be designed and tested in real risk situations. After preliminary calibration, volunteers can be invited to play, to reuse these elements, to build and to simulate their own risk scenarios.

Second, there is the need to develop a sustainable mobilization of the crowds, so as to promote a "collective action" approach to large and systemic risks (T. Maillart, private discussions). While the first proposed approach to complex risks management might interest risk researchers and professionals, its democratized adoption by users of very different backgrounds, socio-economic horizons, age classes and cultures is critical to gather and to organize scattered information, in order to address large scale scenarios. To ensure sustainable mobilization of large populations of users, focus on intrinsic motivation is key. It will be necessary to explore the factors of motivation (hedonic pleasure and personal interests) and their relative proportion from their contribution behaviors. Two kinds of behaviors are expected: in their personal sphere of interests, many individuals will gather and submit the necessary information to document and verify scenarios, while others will rather focus on technical challenges for the pleasure of making a nice design that works. Progressive migration from the first to the second category becomes a proxy of internalization of knowledge and skills by users. Intrinsic motivation ought to drive also individual efforts towards most relevant risk scenarios. As a consequence, having a large number of contributors is the assurance of more accurate design, of better testing and of increased validity. By having many people contribute similar scenarios (or pieces of scenarios), it will be possible to derive quantitative metrics out of qualitative contributions.

Third, it is necessary to develop crowd sourcing to improve the perception of regime shifts and systemic crises. There is always a large part of subjectivity in the way people perceive risks, which are complex, uncertain or even ambiguous. Such biases are likely to emerge as more individuals with various backgrounds and interests will join and contribute to the simulation platform, and therefore, must be considered. In fact, the possibility to capture human perception biases regarding risks at large scales should rather be considered as an opportunity to understand the revealed preferences that, by self-fulfilling prophecies or reflexivity, condition the choices of society. Crowd sourcing is expected to reveal and address idiosyncratic perception biases and further extract systematic ones among large populations. Finally, with contributors coming from various cultural background, differences in the perception of risks should be empirically measured at large scales.

The simulation tools of the "crisis flight simulator" for resilience build-up should be extraordinarily useful for
- (i) scientific synthesis of different fields in a coherent framework,
- (ii) the training of decision makers who do not realize the unintended consequences of their decisions (many of whom are negative and often with enormously bad consequences) and
- (iii) the education of the public, of citizens and of students to be informed as well as to help them direct policy by voting in an informed way.

Different institutions and companies have developed initiatives that have some relationship but are in general much more limited than the presently proposed vision of "crisis flight simulators". One can mention the Japanese Earth Simulator (http://www.jamstec.go.jp/esc/index.en.html, http://en.wikipedia.org/wiki/Earth_Simulator), the Sentient World White Paper (http://www.scribd.com/doc/25656152/Sentient-World-Simulation), Google.org (http://www.google.org/) that utilizes "collective action", Gapminder (http://www.gapminder.org/) for monitoring and visualizing various indices and others.

**6.3 Resilience by multi-variable measurement and prediction**

**6.3.1 Multi-variable measurement of resilience**

In Section 3, it was demonstrated that resilience can be seen as one of the indirect measures of stress used in social sciences. Considering a problem from a different angle, the resilience of a system, i.e., its ability to cope with stress, and its measurement can be improved by taking into account:

1) the multidimensionality of resilience, as the development of a system can be motivated by several goals (subsection 5.2);

2) complementary (preferably direct) dynamical measures of:
- stressors, to which the system is sensitive (e.g. risk measures are used in a probabilistic approach),
- stress, developing within the system (e.g. crash hazard rate),
- costs and efficiency of managerial actions.

As a system is subjected to the influence of numerous factors, which have different effects and are interconnected, it is important that the measurement of resilience would be based not on a single characteristic but include an ensemble of them. It would be very useful to track the dynamics of different stressors and their influence on the stress reaction of the system, as well as monitor how managerial actions affect both of them. Armed with this type of quantitative data, decision makers will be able to better understand the regime in which the system is functioning. They will be able to identify the true source of change in the stress level of the system. The origin of change may include some beneficial dynamics of a stressor, managerial actions, and/or the adaptation of the system to changing conditions. Decision makers may then be able to develop better policy, based on a risk-benefit analysis.

Despite existing limitations, especially in systems that include the "human factor" (see subsection 6.1), theoretical and empirical findings suggest that such a complex quantitative approach to resilience is not only possible but, in many cases, can be enhanced by the development of a predicting capacity.

The next subsection 6.3.2 proposes a more systematic classification of the type of stressors. Then, subsection 6.3.3 builds on the endogenous nature of many crises to suggest the most ambitious approach yet discussed here, namely the "time@risk" approach based on the monitoring of precursors towards the prediction of financial and economic crises. This is nothing but the operational implementation of the famous maxim "Gouverner, c'est prevoir" (governing is predicting) by Emile de Girardin.

### 6.3.2 Analysis of stressor types (exogenous versus endogenous and their interplay)

(1) Stressors can come from external sources and the environment, beyond the direct control of the system. Some are knowable, quantifiable, in the possible losses and their frequencies. This is the favorable situations where counter measures can be build to prepare for the possible losses and to catalyze recovery, using the dynamical framework described in sections 3.2 and 3.3. Considering external stressors, responsible managers and decision makers should also consider the real surprises, such as in the Knightian uncertainty of unknown unknowns popularized by Taleb (2007)'s "black swans". Then, resilience can only be attained with the interplay between, as already said, (i) individual strength and adaptation, (ii) cohesion of the social group as well as (iii) a balance between clear top-down vision that does not exclude the empowerment of individuals at the "bottom" to be able to inform the top and act decisively when needed.

(2) Stressors are also often of an endogenous nature, even if exogenous influences and fluctuating perturbations are always present in out-of-equilibrium open "living" systems. By endogenous, we mean that there is a progressive evolution and maturation of internal interactions between constitutive elements that may give rise to surprising large-scale collective changes. Mathematically, the theory of bifurcations describes well the sudden change of regime from one state or attractor to another one or to a set of other competing attractors upon the small variation of a so-called control parameter. In the bifurcation theory applied to dynamical systems, the fundamental reduction theorem states that bifurcations between states can only occur through a limited number of ways that are known and classified (Thom, 1989; Guckenheimer and Holmes, 1983; Manoel, and Stewart, 2000; Kuznetsov, 2004) and under the change of a small number of (most likely, one) control parameters. Of course, what is the control parameter relevant for a given transition is not known in general but the knowledge that this is the case empowers the decision maker to realize that a given crisis may have a "simple" set of mechanisms after all, whose understanding may be used to track the transition. More precisely, according to this view, it is possible to develop advanced diagnostics of an incoming crisis and invest in techniques to identify precursors. As a corollary, resilience involves precautionary actions that address the observed internal changes. More ambitiously, managers should consider the possibility to change the course and steer the system away from the trouble that is progressively announced by the precursors. In this vein, we claim that many, if not most catastrophes, occur as a surprise because stakeholders and managers have ignored. either by lack of knowledge, insufficient commitment or on purpose, the telling signs of the incoming crisis.

### 6.3.3 Resilience by advanced diagnostics and precautionary actions in finance and economics: the "time@risk" approach

Imagine you had advanced warning signs (and that you listened to them) about the future occurrence of an adverse shock to your firm. Imagine that you could have access to precursory signs of diseases not yet symptomatic in your body (as is the dream of Proteomics). Imagine you could rely on an indicator diagnosing the existence of a financial bubble and indicating the probable time of its burst (as we are developing at the ETH Zurich Observatory of Financial Crises: www.er.ethz.ch/fco/). Imagine that these advanced signs would be revealed years in advance. With this kind of information, you could prepare, you could reflect on what is not working and what could be improved or changed. You could start a process towards building stronger resilience, catalyzed by the knowledge of the nature and severity of the stressors forecasted to come. In contrast to ignorance or complacency, advanced diagnostics could revolutionize risk management by pushing us into action to build defenses. A working advanced diagnostic system would not be static, but would provide continuous updates on possible scenarios and their probabilistic weights, so that a culture of preparedness and adaptation be promoted. This corresponds to exploiting the concept elaborated in section 4 concerning the coevolution of systems and their stressors. Here, we go one step further by suggesting that forecasting the occurrence of crises promotes the evolution of the system towards a higher level of resilience that could not be achieved even by evolution (which is backward looking). Advanced diagnostics of crises constitutes the next level of evolution for cognizant creatures who use advanced scientific tools to forecast their future.

To be concrete, we describe how this system, which we refer to as the "time@risk" approach, would look like when targeting financial and economic instabilities. Here, the outstanding challenge is to develop predictions of systemic risk and global financial instabilities that have emerged as leading concerns in modern economies and with globalization. As Einstein said: "Problems cannot be solved by the same level of thinking that created them." Therefore, a truly interdisciplinary approach to the diagnostic of such crises is required. By leveraging on expertise in Economics, Mathematics, Statistical Physics and Computer Science, a novel integrated and network-oriented approach can be brought to bear on the issue. This would require providing
1. a theoretical framework to measure systemic risk in global financial market and financial networks;
2. an ICT collaborative platform for monitoring global systemic risk;
3. algorithms and models to forecast and visualize interactively possible future scenarios.

Consider the example of a financial crash, such as "black monday" 19 October 1987 mentioned in section 4.4.1. A sum of evidences suggests that it did not come out the blue. Postmortem analysis of many financial crashes shows the development of a kind of standard scenario, as documented for instance by Kindleberger (2005) and Sornette (2003). A financial crash is the result of increasing financial leverage developing together with social herding and the psychology of a "new economy". Specifically, this creates bubbles, and the crashes are nothing but the termination and burst of the bubbles. Using the concept of stress developed throughout the present essay, this endogenous maturation of the financial system towards an instability can be quantified by the excess super-exponential accelerating bubble price. This excess growing price can be used as a *direct measure of the level of stress* increasing within the system. This can be shown via the theoretical linkage between the "crash hazard rate" and the excess price (Johansen et al., 1999; 2000; Yan et al., 2012).

Other *early warning stress signals and diagnostics* for the upcoming transition into the major regime shifts associated with crises include, as reported by (Sornette, 2002; 2004; Dakos et al., 2008; Scheffer, 2009),
(i) a slowing down of the recovery from perturbations,
(ii) increasing or decreasing autocorrelations,
(iii) increasing variance of endogenous fluctuations,
(iv) appearance of flickering and stochastic resonance, and other noise amplification effects (Harras et al., 2012),
(v) increasing spatial coherence, and singular behavior of metrics revealing positive feedbacks (Sammis and Sornette, 2002; Johansen and Sornette, 2010).

This is a very general problem and, in principle, the "time@risk" approach can be extended to various domains of application. The corresponding "time@risk" platform should ideally
(a) signal the possible occurrence of a crisis;
(b) provide insights to adopt the appropriate policy measures; and
(c) allow evaluating future scenarios according to the chosen policy.

The development of a framework for a computational forecasting infrastructure must necessarily combine modeling the relevant entangled networks with empirical analysis and validation of the models. Finally, there is a need to craft the tools into an interactive platform. Therefore, the objectives of the "time@risk" approach can be stated as follows.s
1: Provide novel indicators and methods to estimate the origin and dynamics of systemic risk and forecast probability of systemic crises.
2: Develop agent-based models of the interacting networks which (a) are suitable to be validated, and (b) allow to compute indicators of systemic risk.
3: Validate the models with empirical data.
4: Develop a measurement platform in which it is possible to
(a) load and share relevant data about the involved institutions and their relations,
(b) produce topical maps of interacting networks,
(c) detect the propagation of distress, and
(d) perform simulations, scenario analysis, and systemic risk estimation.

This is an ambitious and risky approach. One should be aware of the risks and difficulties in the development of such a computational forecasting framework. For this reason, tasks should be developed both at empirical and modeling levels and with resources including a collaborative team of experts in an interdisciplinary atmosphere, forecasting technologies combined with the science of networks in order to validate the results obtained. In this way, the following insights can be implemented.
1) In contrast with a majority view of the current understanding, the global industrial, economic, financial and ecological systems are complex in which (a) micro and macro behavior can be dramatically different, (b) density and heterogeneity of the links as well as the whole topology (clusters, cycles and other patterns) may play a role on the (in)stability of the system and (c) time evolution is crucial for spillover effects and externalities to cascade across the system. In this context, equilibrium approaches deliver useful but insufficient and sometimes fundamentally misleading and dangerous insights.

2) It is useful to develop an integrated micro-macro approach including an analysis of a mesoscopic scale in which the system under study is seen as a network of different sectors (e.g. business lines such as commercial banks, investment banks, mutual funds, insurance companies, etc.) with a varying degree of interdependence among them.

3) One can leverage the deep knowledge recently gained by the complex networks community about failure cascades (Buldyrev et al., 2010) and contagion in networks,

4) It is necessary to go beyond the idea, dominant for long times, that big crises need big shocks and offer quantitative understanding of endogenous mechanisms of onset and amplification of crises. In this view, systemic risk is fundamentally different and possibly at odds with individual risk (e.g Morris and Shin 2008, Brunnermeier 2009). In particular, local shocks can also have systemic repercussions (Delli Gatti et al. 2005, Iori et al. 2008; Battiston et al 2007; Sieczka, Sornette and Holyst, 2011).

In the economic and financial applications, the list can be enhanced by the following objectives.

5) A necessary goal is to challenge the mainstream economics vision that more links (and thus interdependence) make always the economy more stable (Allen and Gale, 2000; Shiller, 2004; 2008; Merton and Bodie, 2005). Unfortunately, under some not so infrequent circumstances, financial integration may increase systemic risk (Lorenz and Battiston 2008; Battiston et al 2009). More generally, it has been shown that stronger coupling leads to increased risks of synchronization and to the occurrence of system-wide catastrophes (Sornette, 1994; Osorio et al., 2010). Such events have been termed "dragon-kings" to emphasize their special impact and the specific generating mechanisms (Sornette, 2009; Sornette and Ouillon, 2012).

6) A promising approach is to combine Minsky (1982)'s view, currently under re-evaluation, of an endogenous build-up of financial fragility in the economy with a network approach. As a result, the extent of the systemic repercussions at the Minsky moment depends not only on the distribution of fragility across the agents but also on the structure of their network of mutual financial exposures.

7) It is important to complement the panorama of projects trying to identify precursors of crises from stock prices dynamics, by focusing instead on the network of exposures among financial institutions which play a crucial role in the spreading out financial distress, both in the Money Market (e.g., interbank, Repo, and so on, with maturity < 1 year), in the Capital Market (e.g., bonds, long-term loans, etc. > 1 year) and possibly in the OTC derivatives market.

## 7- CONCLUDING REMARKS

Ideally, an individual, a group or a society would like to be optimized fully for the present, enjoying now the comfort resulting from past achievements and investments while, at the same time, be prepared for the inevitable future stressors that are difficult to foresee. The concept of resilience embodies the quest towards the ability to sustain shocks, be they externally or internally generated or both, to suffer from these shocks as little as possible, for the shortest time possible, and to recover with the full functionalities that existed before the perturbation. Building up resilience is, like risk management, confronted with the eternal conflict between the long-term benefits and the short-term costs. Indeed, building up resilience is costly, as it swallows resources that would otherwise be directed towards optimal present output. And like in risk management, the benefits are visible only when a serious crisis hits the system, which sometimes occur only over time scales of decades. The level of efforts towards resilience can thus be seen to be fundamentally anchored in a kind of philosophical perspective of one's personal life for the individual, or a choice of culture or of society for the larger group. Building up resilience can ultimately be seen as a problem of decision making in the face of conflicting evidence and goals as well as limited strengths in the presence of a complex stochastic environment, with all its complexity and entanglement with all other aspects of life and society. It is a balance between the present versus the future, between commitment for costly investments versus present enjoyments. Yukalov and Sornette (2012) have recently shown that self-organization in complex systems can be treated as decision making (as it is performed by humans) and, vice versa, decision making is nothing but a kind of self-organization in the decision maker nervous systems. Framing the build-up of resilience as a dynamical and continuous decision making process offers novel perspectives, which beg to be explored, based on the bridge between complex pattern formation and evolutionary emergence of novel properties.